# Non-invasive assessment of the spatial and temporal distributions of interstitial fluid pressure, fluid velocity and fluid flow in cancers *in vivo*

Md Tauhidul Islam, Ennio Tasciotti, Raffaella Righetti*

*Abstract*—Interstitial fluid pressure (IFP), interstitial fluid velocity (IFV), interstitial permeability (IP) and vascular permeability (VP) are cancer mechanopathological parameters of great clinical significance. To date, there is a lack of non-invasive techniques that can be used to estimate these parameters *in vivo*. In this study, we designed and tested new ultrasound poroelastography methods capable of estimating the magnitude and spatial distribution of fluid pressure, fluid velocity and fluid flow inside tumors. We theoretically proved that fluid pressure, velocity and flow estimated using poroelastography from a tumor under creep compression are directly related to the underlying IFP, IFV and fluid flow, respectively, differing only in peak values. We also proved that, from the spatial distribution of the fluid pressure estimated using poroelastography, it is possible to derive: the parameter $\alpha$, which quantifies the spatial distribution of the IFP; the ratio between VP and IP and the ratio between the peak IFP and effective vascular pressure in the tumor. Finally, we demonstrated that axial strain time constant (TC) elastograms are directly related to VP and IP in tumors. Our techniques were validated using finite element and ultrasound simulations, while experiments on a human breast cancer animal model were used to show the feasibility of these methods *in vivo*.

## I. INTRODUCTION

The cancer mechanical microenvironment plays an important role in the tumor's growth, invasion and metastasis [1]–[3]. The main components of the cancer mechanical microenvironment are the interstitial fluid pressure (IFP) and the solid stress [2], [4]. There are a number of factors that contribute to the elevated IFP inside a tumor. These factors include blood-vessel leakiness, lymph-vessel abnormalities, interstitial fibrosis and a contraction of the interstitial space mediated by stromal fibroblasts [5], while the uncontrolled proliferation of the cancer cells and the electrostatic force among the negatively charged hyaluronan chains are the main causes for the development of the solid stress inside the tumor [4].

IFP is a parameter of great clinical significance. IFP inside cancers has been identified as one of the major barriers to

This work was supported in part by the U.S. Department of Defense under grant W81XWH-18-1-0544 (BC171600).

Md Tauhidul Islam and *R. Righetti are with the Department of Electrical and Computer Engineering, Texas A&M University, College Station, TX 77843 USA (e-mail: righetti@ece. tamu.edu), Ennio Tasciotti is with Center of Biomimetic Medicine, Houston Methodist Research Institute, 6670 Bertner Avenue, Houston, TX 77030, USA.

cancer treatments [5], [6]. In chemo- and immune-therapy, interstitial fluid flow from the center to the periphery of the tumor induced by the IFP prevents drug molecules to reach the core of the tumor, thus affecting the efficacy and uniformity of drug diffusion in the cancer mass [7], [8]. Interstitial hypertension caused by IFP can cause failure to radiation therapy by increasing cell clonogenicity and VEGF-A expression in tumor tissue [9]. Interstitial fluid flow due to IFP may also promote metastasis, by increasing the shear stress acting on the cancer cells and forcing them to move toward the lymphatic system adjacent to the solid tumor [7], [10]. Due to elevated IFP in cancers, the lymphatic channels inside the tumor are hampered [10] reducing the outflow of fluids from the tumor. All the above phenomena created because of elevated IFP inside the tumor expedite cancer progression, metastasis and reduce therapeutic success. Moreover, IFP has been proved to be a valuable diagnostic and prognostic marker for many types of cancers [11], [12]. Therefore, an imaging method capable of accurately and non-invasively assess IFP in tumors would be of great clinical importance, but it has not been realized yet.

In the past years, several models have been proposed to describe and predict the behavior of IFP in tumors. It has been demonstrated that IFP ($P_i$) inside a spherical tumor can be written using the following expression [13]–[15]:

$$P_i(R) = P_e\left(1 - \frac{\sinh\left(\alpha\frac{R}{a}\right)}{\frac{R}{a}\sinh\alpha}\right), \quad \text{where} \quad \alpha = a\sqrt{\frac{L_p}{k}\frac{S}{V}}, \tag{1}$$

where $P_e$ is the effective vascular pressure. From this equation, it is clear that the spatial distribution of the IFP in the tumor is controlled by the parameter $\alpha$, which is related to the ratio of VP and IP in the tumor [15]. The value of $\alpha$ is clinically important since the spatial distributions of IFP and IFV and ratio between VP and IP can be used for diagnosis of cancers and improve drug delivery therapies [7].

IFV is another important parameter for targeted delivery therapies [7]. IFV is created by the gradient of the IFP from the center to the periphery of the tumor. High IFV inside the tumor is directly related to high intratumor convection, which may facilitate distribution and penetration of drugs throughout the entire tumor [7]. On the other hand, high values of IFV at the periphery of the tumor increase the convective loss of the therapeutics and supply growth factors (VEGF-A and VEGF-C) to surrounding tissues [7], which, in turn, increases the rate of neo-angiogenesis. Therefore, estimation of IFV inside tumors can be useful in drug delivery and for determination of the cancer stage.

IP and VP also carry important information about the tumor mechanopathology. When a solid tumor grows, it initially makes use of existing vasculature but then requires angiogenesis. The newly formed blood vessels are leaky, highly irregular and tortuous [16], [17]. This process increases the VP of the tumor and decreases its perfusion. The IP of the tumor differs from the one of a normal tissue because the proliferation rate of cancer cells results in increased cellular density and collagen deposition in the tumor interstitium [7], [15]. VP and IP modulate drug diffusion and accumulation at the tumor, since they affect convection and consolidation times of the molecules inside the cancer mass [15]. Moreover, knowledge of VP inside the tumor can affect the choice of a treatment to be administered. Vascular normalization treatments are more effective in tumors with high VP, whereas stress normalization treatments are more effective in tumors with low VP [18], [19].

In the past few decades, several invasive techniques have been proposed to estimate the IFP in tumors. These include: transducer-tipped catheter with a precision glide needle [20], wick catheter [21], modified wick technique (wick-in-needle technique) [22], [23], servo-micropipette [24] and subcutaneous capsule implantation for $4 - 6$ weeks [25]. Dynamic contrast-enhanced magnetic resonance imaging (DCE-MRI) has been explored in a small number of studies as a possible tool to estimate IFP/IFV in tissues [26]–[29]. However, values of IFP obtained from these studies show a weak correlation with actual IFP values inside the tumor [27], [28], [30]. Given their clinical importance, IFP and IFV have been extensively investigated both theoretically and experimentally [7], [15], [31]. The spatial parameter of the IFP $\alpha$ has been studied in Refs. [15], [32] and estimated in Ref. [33] using least square fit on the spatial distribution of IFP. However, similarly to IFP and IFV, $\alpha$ can only be experimentally assessed using invasive techniques or expensive contrast-based imaging techniques.

Ultrasound poroelastography is a cost-effective, non-invasive imaging modality that can be used to measure time-dependent strain fields generated in a tissue in response to an externally applied stress [34], [35]. The motivation behind poroelastography is that underlying pathologies such as cancers and lymphedema affect the established fluid pressure gradients and, as such, the induced strains. Fluid pressure created by the applied tissue compression is an important parameter in poroelastography, which affects the resulting displacements, stresses and strains inside the tissue [35]. In this paper, we provide a direct link between the parameters obtained from a poroelastography experiment and the underlying IFP, IFV, fluid flow, VP and IP in the tumor. We first developed analytical formulations to estimate the fluid pressure, fluid velocity, fluid flow, $\alpha$, ratio between VP and IP and ratio between peak IFP and effective vascular pressure from experimental poroelastography data. Then, we mathematically proved that: 1) the spatial parameter of the fluid pressure induced by the applied stress is identical to the spatial parameter $\alpha$ of the IFP; 2) the fluid pressure, fluid velocity and fluid flow estimated using ultrasound poroelastography are the weighted versions of the IFP, IFV and the fluid flow inside the tumor, respectively; and 3) the ratio between VP and IP and the ratio between the peak IFP and effective vascular pressure in the tumor can be estimated from value of $\alpha$. The theoretical models were validated using finite element and ultrasound simulations while the clinical feasibility of the proposed methods was demonstrated in a human triple negative breast cancer animal model before and after treatment.

## II. THEORIES BEHIND THE PROPOSED IMAGING TECHNIQUES

The developed theories for estimation of fluid pressure, fluid velocity, fluid flow, $\alpha$, ratio between VP and IP and ratio between peak IFP and effective vascular pressure in a poroelastic material subjected to a creep compression (i.e., poroelastography experiment) are based on two assumptions:

1) Tumor and normal tissue are poroelastic materials, and they can be modeled as biphasic materials, containing two distinct phases, i.e., a solid phase and a fluid phase.
2) Tumors are of spherical shape and biphasic theory in spherical coordinates can be applied to analyze their poroelastic response to external compression.

The analytical formulations for the estimation of the different parameters are given below.

### A. Fluid pressure

The fluid pressure inside the tumor at radial position $R$ and time $t$ can be written as

$$p(R,t) = -K(\epsilon(R,t) - \epsilon(R,\infty)), \qquad (2)$$

where $K$ is the compression modulus of the tumor and $\epsilon$ is the volumetric strain inside the tumor estimated from poroelastography experiments.

### B. Ratio between vascular permeability and interstitial permeability and parameter $\alpha$

By solving the continuity equation of pore fluid in a poroelastic material, when the fluid pressure is zero at the periphery of a tumor, the equation for the fluid pressure inside the tumor can be written as (see section 1 of the supplementary information [1])

$$p(R) = \Psi \left(1 - \frac{\sinh\left(\alpha \frac{R}{a}\right)}{\frac{R}{a} \sinh \alpha}\right), \quad \text{where} \quad \alpha = a\sqrt{\frac{L_p}{k}\frac{S}{V}}. \qquad (3)$$

Here $\Psi$ is a constant related to the peak fluid pressure $P_0$ as $P_0 = \Psi(1 - \alpha \operatorname{cosech}(\alpha))$ and $a$ is the radius of the tumor. $L_p$ is the VP, $k$ is the IP and $S/V$ is the surface area to volume ratio of the capillary walls inside the tumor. This equation resembles the equation of IFP (eq. (1)), originally derived in Refs. [13]–[15], except for the constant $\Psi$. In eq. (3), the constant $\Psi$ depends on the applied pressure in the poroelastography experiments, whereas the constant $P_e$ in the equation of IFP derived in Refs. [13]–[15] can be written as $P_e = P_B - \sigma_s(\pi_B - \pi_i)$, where $P_B$ is the microvascular pressure, and $\pi_B$ and $\pi_i$ are the plasma osmotic pressure and the interstitial osmotic pressure, respectively.

---

[1]supplementary information is available in the supplementary files/ multimedia tab

The parameter $\alpha$ can be estimated by fitting the fluid pressure data (estimated using eq. (2)) with the theoretical equation of fluid pressure (eq. (3)). Using the value of the radius of the tumor and the surface area to volume ratio of the capillary walls, it is possible to determine the ratio between the VP and IP by knowledge of $\alpha$.

### C. Ratio between peak IFP and effective vascular pressure

The ratio between the peak IFP and effective vascular pressure can be expressed as

$$\frac{P_{i0}}{P_e} = 1 - \alpha\, \text{cosech}(\alpha). \quad (4)$$

Using the estimated value of $\alpha$ in eq. (4), the ratio between the peak IFP and effective vascular pressure can be determined.

### D. Fluid velocity

The permeability-normalized fluid velocity with respect to the solid along the radial direction inside a tumor can be written as

$$v_R(R,t) = -\frac{dp(R,t)}{dR}. \quad (5)$$

As the fluid pressure $p$ is the weighted version of IFP (eq. (3)), the fluid velocity $v_R$ is also the weighted version of the IFV.

### E. Fluid flow

We can write the equation for the fluid flow occurring in the tumor during a poroelastography experiment as

$$w = \frac{\delta \epsilon}{\delta t}. \quad (6)$$

Alternatively, the fluid flow can be expressed as the gradient of the fluid velocity occurring during a poroelastography experiment multiplied by the porosity of the tumor (see section 1 of supplementary information). Similarly, the fluid flow inside the tumor can be computed as the product of the gradient of the IFV and the porosity of the tumor. Therefore, the fluid flow occurring in the tumor during a poroelastography experiment is the fluid flow inside the tumor multiplied by a constant factor.

### F. Axial strain time constant

The axial strain time constant $\tau$ can be defined as [35]

$$\tau = \frac{\Omega}{H_A k} + \frac{1}{H_A \chi}, \chi \approx L_p \frac{S}{V}, \quad (7)$$

where $\Omega$ is a constant, which depends on the Poisson's ratio and radius of the tumor. We see from eq. (7) that $\tau$ is inversely proportional to the values of the IP and VP.

The detail proofs of eqs. (2)-(7) are shown in section 1 of the supplementary information.

### G. Validation of the proposed techniques

Estimation of the fluid pressure (eq. (2)), $\alpha$ (eq. (3)) and the fluid velocity (eq. (5)) using the proposed technique was validated using finite element and ultrasound simulations as reported in section 2 of the supplementary information. Fluid flow and axial strain TC were not directly available from finite element simulation software but were computed by established theories from the literature [36], [37].

## III. Experiments

Mice implanted with triple negative breast cancer were scanned once a week for three subsequent weeks. The cancers were created by injecting cancer cells orthotopically in the mammary fat pad of the mice [38]. *In vivo* experiments were approved by the Houston Methodist Research Institute, Institutional Animal Care and Use Committee (ACUC-approved protocol # AUP-0614-0033). Mice were kept untreated (n=6), or treated with FEPI (Epirubicin alone, n=3), LEPI (liposomes loaded with Epirubicin, n=3) and LEPILOX (liposomes loaded with Epirubicin and conjugated with a targeting anti-LOX antibody on the particle surface, n=3) for three weeks. The dose of each drug was 3 mg/kg body weight once a week. Mice were individually scanned in dedicated ultrasound imaging session. Each acquisition was 5 minutes long, and several RF data were obtained in the scan. For the experiment, mice were anesthetized with isoflurane and kept lying on a thermostat-regulated heating pad to prevent movement and discomfort.

Elastography was carried out using a 38-mm linear array transducer (Sonix RP, Ultrasonix, Richmond, BC, Canada) with a center frequency of 6.6 MHz, 5-14 MHz bandwidth. To achieve a uniform surface geometry and improve focusing, an aqueous ultrasound gel pad (Aquaflex, Parker Laboratories, NJ, USA) was placed between the compressor plate and the tumor mass. A force sensor (Tekscan FlexiForce) was placed between the top surface of the gel pad and the compressor plate to record the applied force during compression. Creep experiments were performed on the animals by applying manual compression, with the duration of each experiment being one minute [39]. The ultrasound RF data acquisition was synchronized to the application of compression. The sampling rate of the data was configured as 0.1 second/sample. The axial and lateral strain data were calculated at a specific time point by summing all the strains (between two consecutive RF frames) from the start to end of that time point [40]. The applied compressions in the creep experiments were inspected using a graphical user interface software purchased with the force sensor. The borders of the cancers were drawn from the *in vivo* axial strain elastograms.

### A. Estimation of axial and lateral displacements and strains

To compute the axial and lateral strain elastograms from ultrasound pre- and post-compressed RF data, the method proposed in Ref. [41] was used.

### B. Estimation of Young's modulus, Poisson's ratio and axial strain TC

For estimating the fluid pressure, estimation of the Young's modulus and Poisson's ratio of the tumor is required, which was performed using the method described in Ref. [39]. Variable projection method proposed in Ref. [42] was used to estimate the TC of the axial strain temporal curve.

*C. Estimation of fluid pressure, velocity and flow*

We used eqs. (2) and (5) to estimate the fluid pressure and velocity from *in vivo* experimental data. We applied a combination of Kalman and non-linear complex diffusion filters on the axial and lateral strains to remove the noise before computing the fluid pressure using them [43]. The reconstructed fluid pressures and fluid velocities reported in this paper are at time point of 10 s. The fluid flow is computed using eq. (6), i.e., by taking the time differential of the volumetric strain with respect to two time points, i.e., $t_1 = 10$ s and $t_2 = 60$ s ($w = \frac{\epsilon_2 - \epsilon_1}{t_2 - t_1}$). The reconstructed fluid pressure, fluid velocity and fluid flow were normalized by dividing them by the applied pressure. Thus, they correspond to 1 kPa applied pressure, which ensured a fair comparison among these parameters in treated and untreated tumors at the different time points.

*D. Estimation of $\alpha$*

To determine the value of $\alpha$ *in vivo*, we fit the fluid pressure estimated from experimental strain data (from the center to a radial direction) with the theoretical equation of the fluid pressure (eq. (3)). We normalized the fluid pressure curve by dividing it by the peak value and then fit eq. (3) onto it by varying both $\Psi$ and $\alpha$. Estimation of $\alpha$ in this manner does not require knowledge of $\Psi$. We used 'Levenberg Marquardt' algorithm in Matlab (Matlab Inc, Natick, MA, USA) as the curve fitting algorithm. A smoothing filter of length 5 pixels was applied on the fluid pressure data before estimating the $\alpha$.

*E. Computation of surface area to volume ratio of the capillary walls inside the tumor*

For computing the surface area to volume ratio of the capillary walls inside the tumor, we used the following equation [44]

$$\frac{S}{V} = 10 f V_t^g. \qquad (8)$$

where $V_t$ is the volume of the tumor, which is computed as $V_t = \frac{4}{3}\pi a^3$, $f = 54.68, g = -0.2021$ [44]. Here, $a$ is in units of mm and $\frac{S}{V}$ is in units of cm$^{-1}$.

*F. Statistical Analysis*

Data in Figs. 3 and 4 are presented as mean ± SD (standard deviation). Matlab (MathWorks Inc., Natick, MA, USA) was used to analyze the data. Statistical significance was determined using the Kruskal-Wallis test.

IV. RESULTS

Fig. 1 shows B-mode images and corresponding reconstructed fluid pressures, fluid velocities, fluid flows and axial strain TC elastograms at three time points (week 1, week 2, week 3) for an untreated tumor. The untreated tumor increased in size with time as appreciable from the B-mode images (A1, B1, C1). The fluid pressures (A2, B2, C2) appeared spatially uniform especially at week 3. At week 1, the fluid velocity was more than 0.02 kPa per pixel in most of the regions inside the tumor. In the second and third weeks, the fluid velocity decreased and reached values close to zero in most of the regions inside the tumor. The fluid flow (A4, B4, C4) reduced with time from the first to the third week and overall exhibited very low values in the second and third weeks.

The axial strain TC (A5, B5, C5) decreased with time in the untreated tumor. As the effect of the IP is typically much smaller than the effect of the VP in most tumors [7], [15], it is reasonable to assume that the reduction in the axial strain TC in the tumors was primarily due to changes in the aggregate modulus or the VP (see eq. (7)). Based on the values of Young's modulus and Poisson's ratio of this tumor in week 1, 2 and 3 (see Table I), the change in aggregate modulus from week 1 to 3 was at most 1.5 folds. However, the overall reduction of axial strain TC from the first to the third week was about three times (90 to 30 s). Therefore, the decrease of the axial strain TC value was mainly due to the increase of VP, consistently with what has been previously reported [45].

B-mode images and reconstructed fluid pressures, fluid velocities, fluid flows, axial strain TC elastograms in a treated tumor are shown in Fig. 2. As a result of the treatment, tumor decreased in size from week 1 to week 3 consistently with what seen in B-mode images (A1, B1, C1). As opposed to the untreated tumor, the fluid pressures (A2, B2, C2) inside the treated tumor were space-dependent in all three weeks. The fluid velocity (A3, B3, C3) increased with time in the treated tumor. The values of fluid velocity also increased with respect to the untreated one. Fluid flow (A4, B4, C4) was seen increasing with time. When compared to the untreated tumors, the value of fluid flow was also higher in the treated tumors.

The axial strain TC values (A5, B5, C5) increased with time in the treated tumor. The average axial strain TC of the treated tumor was found to be around 50 s in the first week and $> 100$ s in the third week while in the untreated tumor the average axial strain TC was in the range $20 - 30$ s (in third week). Since the Young's modulus ($< 40$ kPa) and Poisson's ratio ($0.29 - 0.35$) did not change significantly (see Table I), this increase in axial strain TC in the treated tumor was likely to be attributed to a reduction of vascular permeability.

The bar plots of the fluid velocity (A1), fluid flow (A2), axial strain TC (A3), $\alpha$ (B1), ratio of VP and IP (B2) and ratio of the peak IFP and effective vascular pressure (B3) are shown in Fig. 3. These plots show mean value over standard deviation of different parameters. Overall, based on the mean values, fluid velocity, fluid flow and axial strain TC reduced with time in the untreated tumors and increased with time in the treated tumors.

The mean value of $\alpha$ (B1) increased consistently with time in the untreated tumors whereas it decreased in the treated tumors. The mean value of $\alpha$ was lower for the treated tumors than the untreated ones in the second and third weeks. According to previous literature [7], $\alpha$ typically becomes higher as a cancer progresses and decreases with treatment. The mean ratio between VP and IP in untreated and treated tumors is shown in Fig. 3 (B2). Similar to $\alpha$, the mean ratio between VP and IP was much higher in the case of the untreated tumors than the treated tumors in the second and third weeks. This may indicate a net reduction of VP in treated tumors because of drug administration as the administered drugs were reported to have insignificant/no effect on IP [7]. In treated tumors, the mean ratio between VP and IP decreased with time, whereas it increased in untreated tumors, consistently with previous observations [7]. The average values of the ratio between the peak IFP and the effective vascular pressure at the different time points are shown in Fig. 3 (B3). The mean value of this ratio increased with time inside the untreated tumors, whereas it decreased in the treated tumors. This implies that the IFP increased in untreated tumors, while it decreased in treated tumors if the effective vascular pressure is assumed constant with time inside a tumor [46].

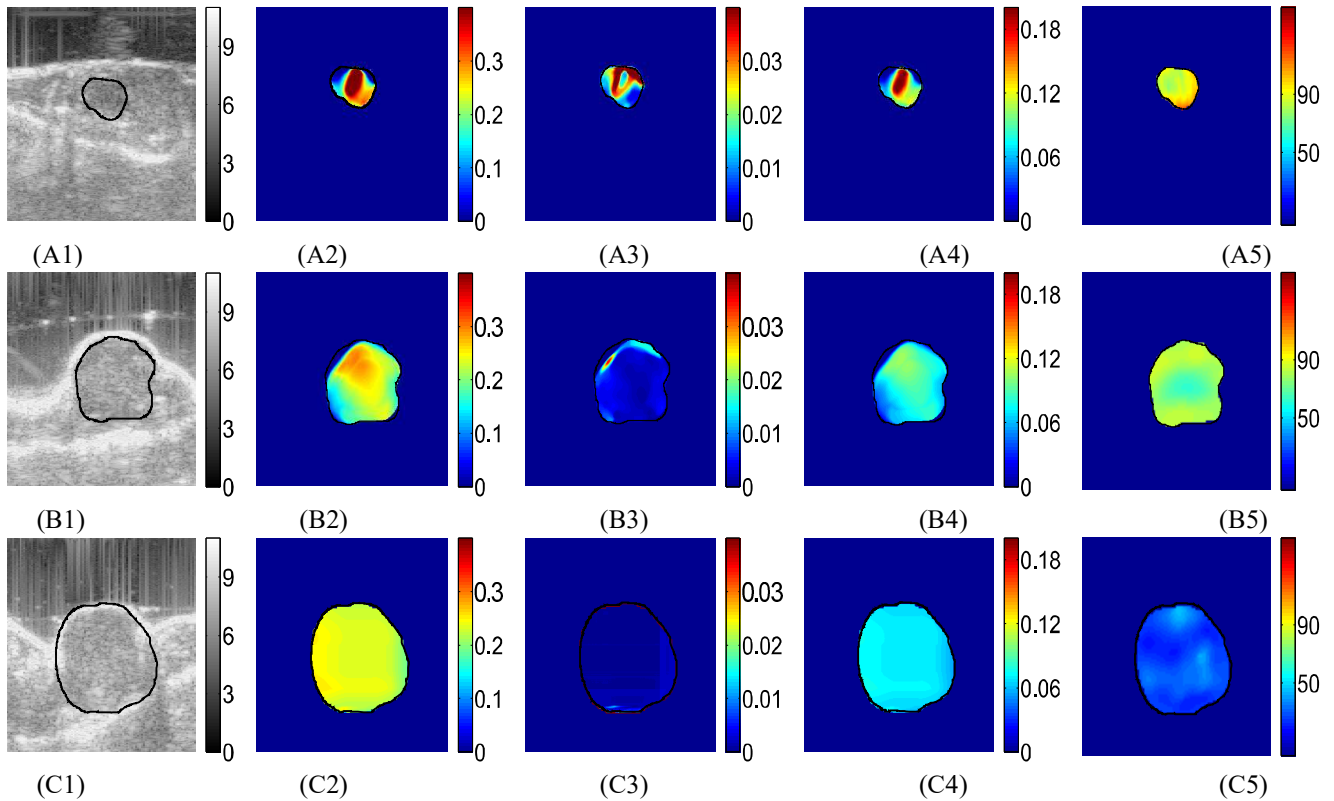

Fig. 1: Ultrasound B-mode images of an untreated tumor at three time points (week 1, week 2 and week 3) are shown in A1, B1 and C1, respectively. Fluid pressures at these time points are shown in A2, B2 and C2. Reconstructed fluid velocities inside the tumor at the three time points are shown in A3, B3 and C3, respectively. Reconstructed fluid flows inside the same tumor at three time points are shown in A4, B4 and C4, respectively. Axial strain TC elastograms of the same tumor estimated from axial strain temporal curve at the three time points are shown in A5, B5 and C5, respectively. The fluid pressures and fluid velocities are in scale of kPa and kPa per pixel, respectively. The length of each pixel is $0.3125$ mm. The fluid flows are in unit of milli-strain s$^{-1}$. The unit of axial strain TC is second. The size of the tumor increased consistently from week 1 to week 3 as seen from the B-mode images. The fluid pressure was very smooth and there was little spatial variation inside the tumor especially at week 2 and week 3. The fluid velocity was very small at week 2 and week 3 for this tumor as seen in (B3) and (C3). The fluid flow and axial strain TC seemed decreasing in value from week 1 to week 3.

In Fig. 4 (A), (B) and (C), we report the Young's modulus and surface area of the tumors and surface area to volume ratio of the capillary walls inside the tumors, respectively. Our results show that Young's modulus and surface area increased and the mean surface area to volume ratio of the capillary walls ($S/V$) reduced with time in case of the untreated tumors, whereas they remained almost constant in the treated ones. Such changes in values of $S/V$ imply that the microvascular density decreases with time in untreated tumors, while it remains almost the same in treated tumors. This observation correlates with results in the literature showing that tumor microvascular density decreases with cancer progression [44].

## V. DISCUSSION

In this paper, we developed methods to image fluid pressure, fluid velocity, axial strain TC and fluid flow inside a tumor using ultrasound poroelastography techniques and demonstrated the existence of a link between the estimated parameters and the underlying tumor mechanopathological parameters. Fluid pressure, fluid velocity and fluid flow estimated using poroelastography can be used to assess the spatial distribution of the actual underlying IFP, IFV and fluid flow parameters and find regions in the tumor where these parameters are higher or lower than average. This information could be of great clinical significance for cancer diagnosis and treatment.

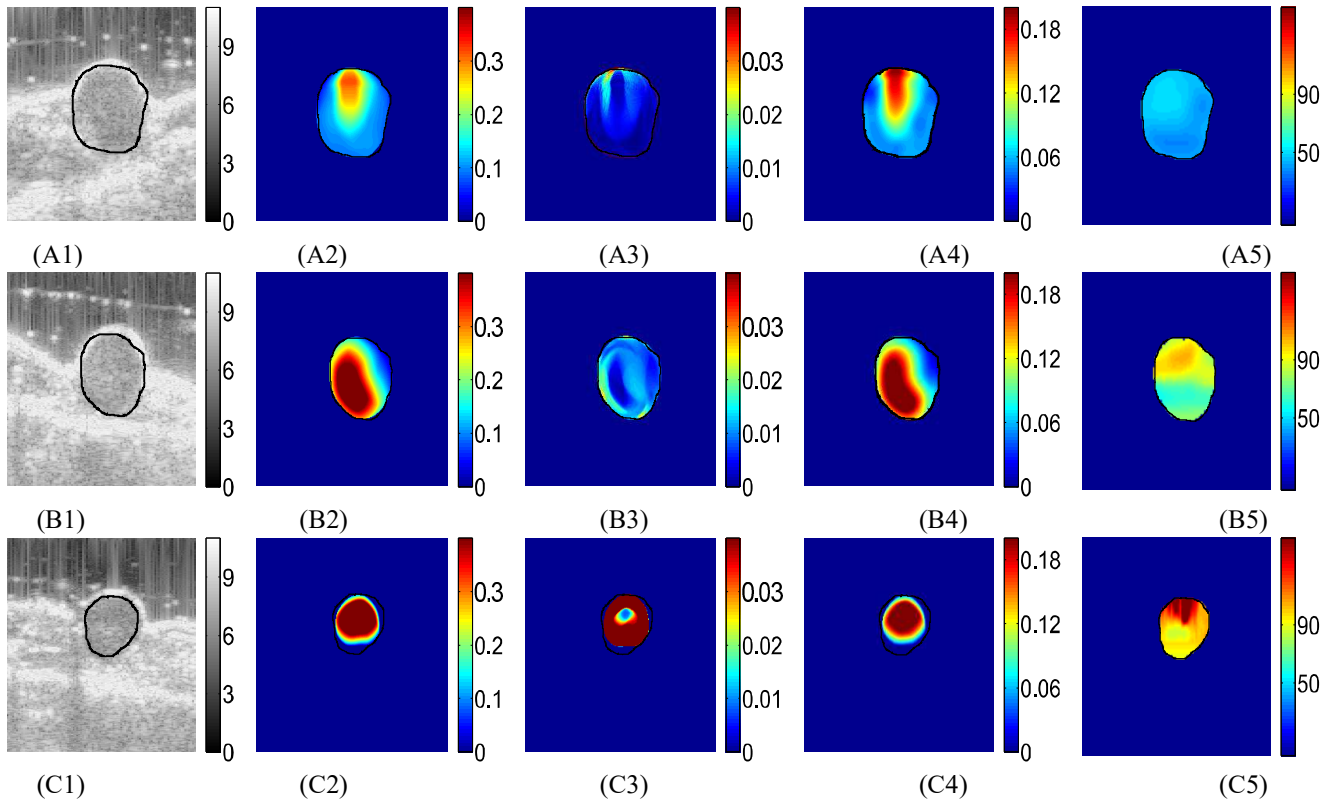

Fig. 2: Ultrasound B-mode images of the first treated tumor at three time points (week 1, week 2 and week 3) are shown in A1, B1 and C1, respectively. Fluid pressures at these time points are shown in A2, B2 and C2. Reconstructed fluid velocities inside the tumor at the three time points are shown in A3, B3 and C3, respectively. Reconstructed fluid flows inside the same tumor at three time points are shown in A4, B4 and C4, respectively. Axial strain TC elastograms of the same tumor estimated from axial strain temporal curve at the three time points are shown in A5, B5 and C5, respectively. The fluid pressures and fluid velocities are in scale of kPa and kPa per pixel, respectively. The length of each pixel is $0.3125$ mm. The fluid flows are in unit of milli-strain s$^{-1}$. The unit of axial strain TC is second. Unlike the untreated tumors, the size of the treated tumor decreased consistently from week 1 to week 3 because of treatment administration. The fluid pressure was spatially dependent in the treated tumor in all three weeks which is in contrast with the untreated tumor. The fluid velocity was higher in case of this treated tumor in comparison to the untreated one. The fluid flow and axial strain TC increased with time in treated tumors, contrarily to the untreated tumors, where these parameters decreased with time.

TABLE I: Young's modulus and Poisson's ratio of treated and untreated tumors at different time points

| Sample | Young's modulus (kPa) | | | Poisson's ratio | | |
|---|---|---|---|---|---|---|
| | Week 1 | Week 2 | Week 3 | Week 1 | Week 2 | Week 3 |
| Untreated tumor #1 | 63.54 | 77.61 | 97.12 | 0.29 | 0.25 | 0.24 |
| Treated tumor #1 | 34.09 | 28.54 | 22.88 | 0.35 | 0.35 | 0.29 |

Several observations can be inferred from the results presented in this paper. In terms of the fluid velocity, we found that fluid velocity was high inside treated tumors. This typically occurs when IFP is low and space-dependent inside the tumor [32]. Due to the treatment, it may be possible that the compressive stress in the tumor interstitium decreased and also that the number of cancer cells and collagen deposition in the tumor reduced (as the Young's modulus of this tumor decreased (see Table I)). A reduction of cell density and collagen content may effectively improve the tortuosity of the matrix, thus resulting in higher fluid velocities [47]. However, in the case of untreated tumors in the third week, the fluid velocity was found to be close to zero everywhere inside the tumor. Such very low fluid velocity ($\approx 0$) typically occurs when the peak value of IFP is spatially constant and high inside the tumor ($\approx P_e$).

Fluid flow is directly related to the outflow compliance of the fluid (and/or drugs) from inside the tumor. In normal tissues, the lymphatic vessels are intact and fully functional, IFP is zero and the fluid flow is normal. In the cancer parenchyma, because of the abnormal lymphatic vessels, contraction of the vasculature due to solid stress and presence of elevated IFP, the fluid flow is hampered and reduces in values [45],

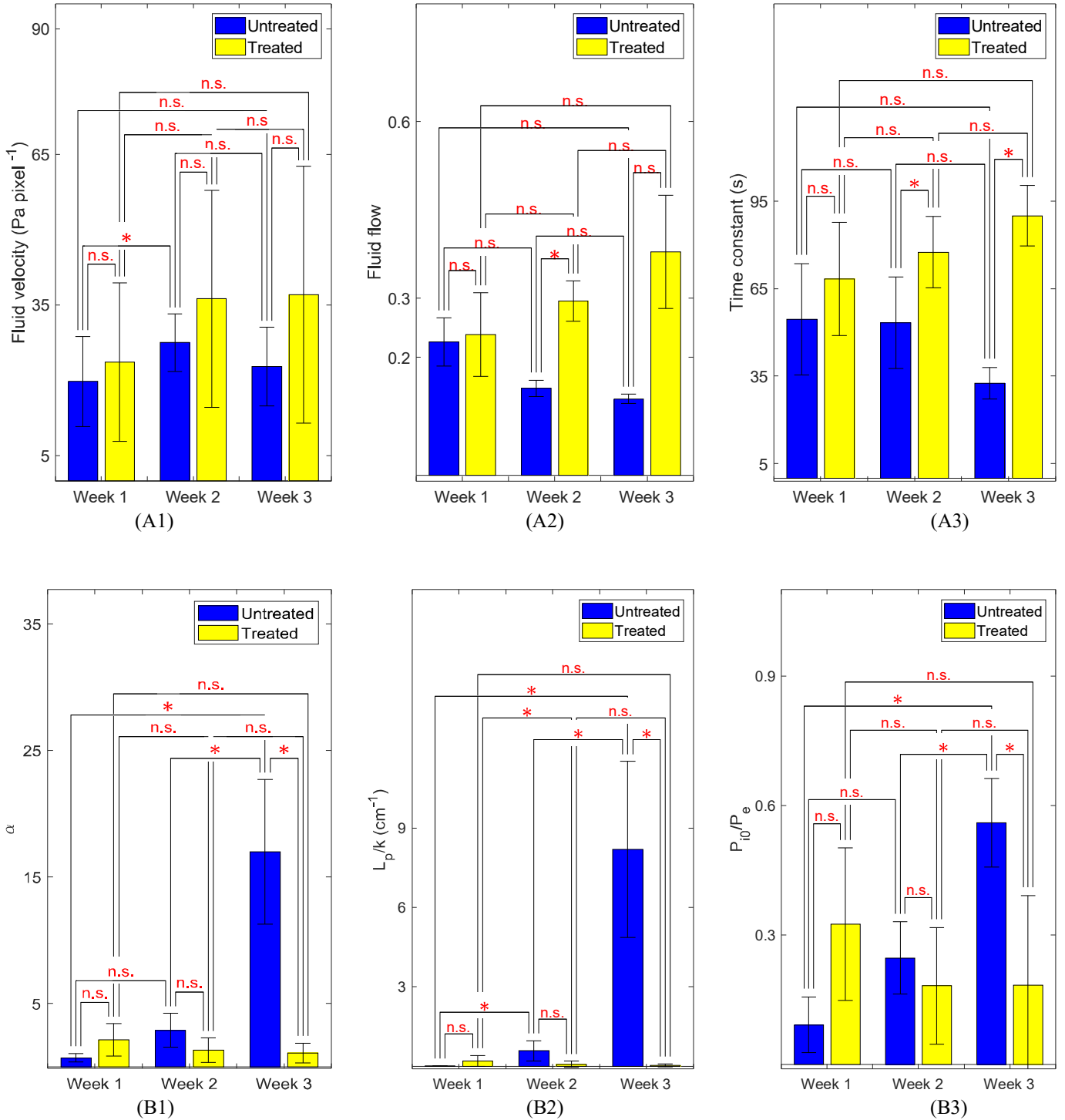

Fig. 3: (A1) Mean fluid velocities in unit of Pa per pixel inside the tumors of the treated and untreated tumors at week 1, week 2 and week 3. (A2) Mean fluid flows inside the treated and untreated tumors at week 1, week 2 and week 3. The unit of fluid flow is milli-strain per second. (A3) Mean axial strain TCs inside the treated and untreated tumors at week 1, week 2 and week 3. (B1) Mean values of $\alpha$ inside the treated and untreated tumors at week 1, week 2 and week 3. (B2) Mean ratios between the vascular and interstitial permeability inside the treated and untreated tumors at week 1, week 2 and week 3. (B3) Mean ratios between the peak IFP and effective vascular pressure inside the treated and untreated tumors at week 1, week 2 and week 3. 'n.s.' means not statistically significant. One and two stars correspond to $p$-value less than $0.05$ and $0.01$, respectively. Based on only mean values, the fluid velocity, axial strain TC and fluid flow decreased with time in untreated tumors and increased with time in the treated ones. On the other hand, the mean value of $\alpha$, ratio between the vascular and interstitial permeability and ratio between the peak IFP and effective vascular pressure increased with time in untreated tumors and decreased with time in the treated ones.

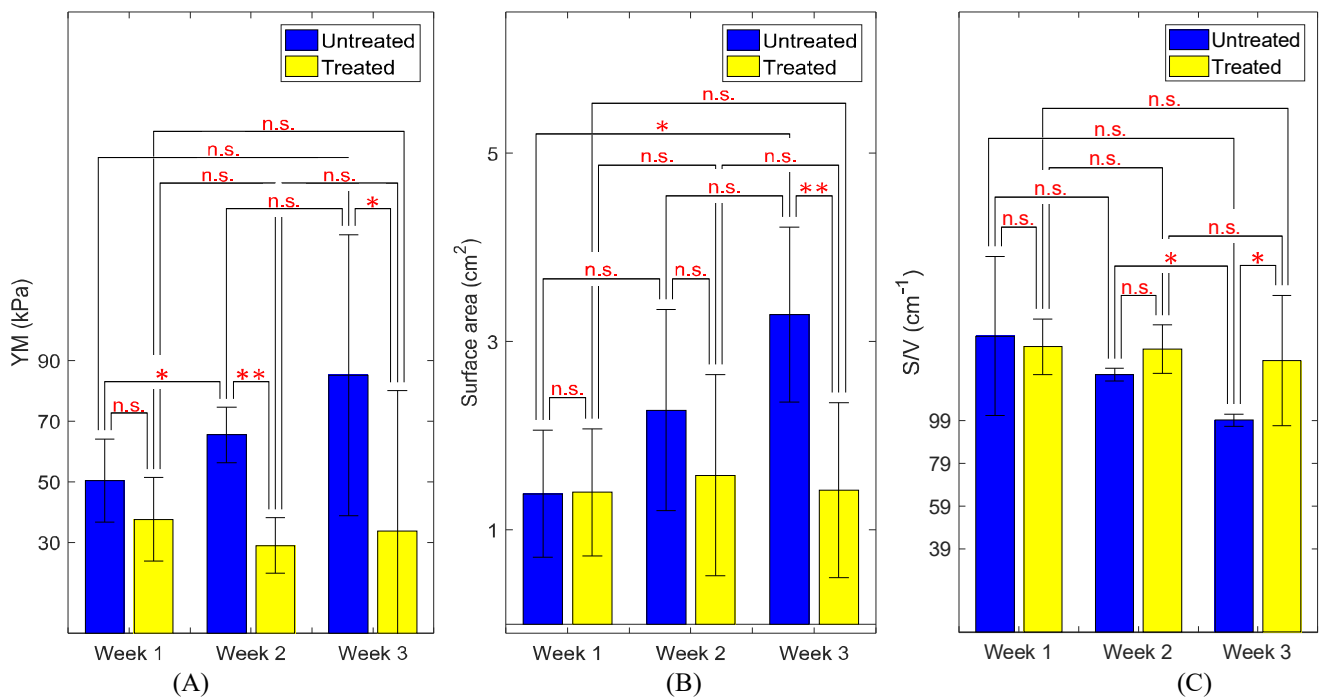

Fig. 4: (A) Young's modulus and (B) surface area of the tumors, (C) Surface area to volume ratio of the capillary walls inside the tumors at three weeks. 'n.s.' means not statistically significant. One and two stars correspond to $p$-value less than $0.05$ and $0.01$, respectively. Based on only mean values, the Young's modulus and surface area of the untreated tumors increased consistently with time, whereas values of these parameters remained almost same in the treated tumors with time. The surface area to volume ratio of the capillary walls decreased in untreated tumors with time, whereas remained almost same in the treated tumors at all time points.

[48]. This phenomenon appears to be confirmed by the *in vivo* experimental results obtained in untreated animals (Fig. 3 (A2)). Knowledge of fluid flow in the tumors has been shown to provide important information for drug delivery therapies as the rate of fluid flow dictates the extravasation of drug molecules from tumor tissue [49].

Another important result of our study relates to the potential significance of axial strain TC images in cancer applications. Using an analytical model of poroelastic response in spherical tumors and knowledge of the aggregate modulus of the tumor, in the future, it could be possible to determine both the IP and VP of a tumor from the estimated axial strain TC elastograms [35]. In addition, if the effect of IP can be assumed to be small in comparison to that of VP inside the tumor [15], with known values of the axial strain TC and the aggregate modulus inside the tumor, we can estimate the VP using models described in Ref. [35].

In this study, we reported a method to assess the parameter $\alpha$, which dictates the spatial distribution of IFP inside a tumor and, as such, can be used in designing drug delivery techniques [7]. In tumors with large $\alpha$, the gradient of the IFP is very small inside the tumor and very high at the periphery. Therefore, most of the drug cannot reach the central portion of the tumor and accumulates at the periphery [7]. If $\alpha$ reduces, the gradient of the IFP (IFV) increases and the drug has a better chance to reach the central portion of the tumor. However, if $\alpha$ becomes too low ($\leq 1$), the IFP becomes very small and the IFV again reduces and the effective drug delivery gets hampered. Therefore, treatments such as vascular normalization should be administered in such a way that $\alpha$ should be close to 5, so that proper drug delivery may be achieved [7]. In tumors with low $\alpha$ (close to 5), it is also possible that the flux of growth factors reaching the draining lymph nodes is decreased due to less fluid flow from the boundary [7], and it could also inhibit lymph node lymphangiogenesis [50]. Lymph node lymphangiogenesis is thought to potentially increase the incidence of lymph node metastasis, by providing additional opportunities for the cells to enter into the lymphatic system. Therefore, a low value of $\alpha$ may reflect in an improvement of the delivery/penetration of therapeutics in tumors, alleviation of peritumoral fluid accumulation, and, at the same time, decrease of the shedding of cancer cells into peritumoral fluids or surrounding tissues.

In the past, methods to estimate $\alpha$ have been proposed but with some limitations. For example, the method proposed in [33] requires values of IFP along the full radius of the tumor to use least square fit on the spatial distribution of IFP. This method produces erroneous results when the IFP is low/zero inside the tumor or when all IFP values are not available along the radius of the tumor. This may be the case, when IFP is measured at discrete points inside the tumor such as in invasive methods involving wick and needle/servo-micropipette [22]–[24]. Our proposed method for estimation of $\alpha$ does not depend on the value of IFP inside the tumor and is able to estimate $\alpha$ even when the IFP is zero inside the tumor.

In this study, we showed that it is possible to estimate the ratio between VP and IP and the ratio between the peak IFP and effective vascular pressure inside the tumor using poroelastography. The ratio between VP and IP may be used as a marker of the efficacy of certain cancer treatments. As an example, if vascular normalization treatment is used and is effective, this ratio may be expected to decrease [45]. Change in the value of the ratio between peak IFP and effective vascular pressure can be a direct indicator of the change in the peak value of IFP inside the tumor, when the effective vascular pressure (approximately equal to microvascular pressure [7]) remains constant over time [46]. Along with all the above-mentioned parameters, we also plotted the surface area to volume ratio of the capillary walls, surface area and Young's modulus of the treated and untreated tumors to further characterize the *in vivo* results.

The main limitation of the proposed approach is that the reconstructed interstitial fluid pressure, fluid velocity and fluid flow do not represent the actual values of IFP, IFV and fluid flow inside the cancer tumor. Rather, they represent their weighted values. However, based on the studies presented in Refs. [7], [45], the spatial and temporal distributions of these parameters can oftentimes be sufficient for diagnosis and treatment planning of cancers even if the actual values may be unknown.

## VI. CONCLUSIONS

In this paper, we applied poroelastography-based methods to determine fluid pressure, fluid velocity, $\alpha$, axial strain TC and fluid flow inside a tumor as a result of an externally applied compression. We proved mathematically that these estimated parameters are related to the actual IFP, IFV, fluid flow, IP and VP inside the tumors. Thus, the estimated parameters using poroelastography may be of great clinical significance for diagnosis, prognosis and treatment of cancers.

# Non-invasive assessment of the spatial and temporal distributions of interstitial fluid pressure, fluid velocity and fluid flow in cancers *in vivo* Supplementary information

Islam et al.

## 1. Detail proofs of the formulations for estimation of the proposed parameters

Based on the theory of Eshelby, the applied uniaxial stress ($\sigma_a$) from the top of the sample in an elastography experiment (see Fig. S1) is inflicted over the full outer surface of the spherical inclusion (tumor) (1). Therefore, when we are interested in the analysis of strains, fluid pressure inside the tumor, the problem can be thought as one of a poroelastic sphere under a uniform compressive/volumetric stress ($\sigma$) over its outer surface. The stress $\sigma$ over the sphere can be computed using Eshelby's theory from the applied stress $\sigma_a$, geometry and Young's modulus and Poisson's ratio of the tumor. In such case, based on the assumption of smaller fluid pressure outside the tumor (because of higher interstitial permeability in the normal tissue), the volumetric strain and fluid pressure become functions of only $R$ and $t$ inside the poroelastic tumor (2). Consequently, the problem can be simplified from cylindrical coordinate system to spherical coordinate system. Based on that, we provide the formulations for different parameters in spherical coordinate below.

**A. Fluid pressure.** Let us consider a poroelastic sample of volume V, which contains a solid phase of volume $V_s$ and a fluid phase of volume $V_p$. The total volume can be written as the sum of these two volumes.

$$V = V_p + V_s. \quad [1]$$

In the above equation, any isolated pores that the fluid cannot infiltrate are considered part of the solid phase. To construct the constitutive equations, we select the volumetric strains of the sample and its fluid phase as the dynamic variables, i.e., ($\Delta V/V, \Delta V_p/V_p$). The following formulations can be developed based on the theory of the effective stresses, (3, p. 84)

$$\frac{\Delta V}{V} = -\frac{1}{K}(\sigma - \zeta p), \quad \frac{\Delta V_p}{V_p} = -\frac{1}{K_p}(\sigma - \beta p), \quad [2]$$

where $\zeta$ and $\beta$ are the Biot's effective stress coefficient and pore volume effective stress coefficient, respectively, $p$ is the fluid pressure and $K$ and $K_p$ are the drained bulk modulus and pore volume bulk modulus, respectively. It should be noted that the above described relationships are a generalization of Hooke's law.

The remaining of our analysis is based on the assumption of intrinsic incompressibility for both the solid phase and the fluid phase, for which $\zeta = 1$, $\beta = 1$ and $K_p = \infty$.

Based on this assumption, we can write

$$\frac{\Delta V}{V} = -\frac{1}{K}(\sigma - p), \quad \frac{\Delta V_p}{V_p} = 0. \quad [3]$$

The left hand side of the first equation is the volumetric strain. In an axisymmetric setup, the volumetric strain $\epsilon$ can be written as

$$\epsilon = -\frac{\Delta V}{V} = \epsilon_{zz} + 2\epsilon_{rr}. \quad [4]$$

where

$$\epsilon_{zz} = \frac{du_z}{dz}, \quad \epsilon_{rr} = \frac{du_r}{dr}. \quad [5]$$

The radial and axial displacements are denoted using $u_z$ and $u_r$.

Based on the above discussion, the volumetric strain and fluid pressure at any time $t$ inside a spherical tumor are related by

$$\epsilon(R,t) = \frac{1}{K}(\sigma - p(R,t)), \qquad [6]$$

where $K$ is the compression modulus of the tumor. At steady state ($t = \infty$), when the fluid pressure becomes zero, the volumetric strain can be written as

$$\epsilon(R,\infty) = \frac{1}{K}(\sigma). \qquad [7]$$

Therefore, we obtain

$$\epsilon(R,t) - \epsilon(R,\infty) = \frac{1}{K}(-p(R,t)). \qquad [8]$$

The fluid pressure at time $t$ can be written as

$$p(R,t) = -K(\epsilon(R,t) - \epsilon(R,\infty)). \qquad [9]$$

**B. Ratio between vascular permeability and interstitial permeability and parameter $\alpha$.** We can write the following differential equation for the fluid pressure inside a spherical tumor with IP $k$ and VP $L_p$ (4)

$$\frac{1}{H_A}\frac{dp}{dt} + \frac{1}{H_A}\frac{dQ}{dt} + \chi p = k\left(\frac{d^2p}{dR^2} + \frac{2}{R}\frac{dp}{dR}\right), \qquad [10]$$

where $Q$ is an integration constant and depends only on $t$, $H_A$ is the aggregate modulus of the tumor and $\chi$ is the average microfiltration coefficient. Here, $\chi = \chi_V + \chi_L$, with $\chi_V = \frac{L_p S}{V}$ and $\chi_L = \frac{L_{PL} S_L}{V_L}$. $L_p$ and $L_{pL}$ are the permeability of the capillary and lymphatic walls. $\frac{S}{V}$ and $\frac{S_L}{V_L}$ are the surface area to volume ratio of the capillary and lymphatic walls. Based on the values of the permeabilities of capillary and lymphatic walls reported in the literature, $\chi_V >> \chi_L$ (5). This results in $\chi \approx \chi_V$, and the microfiltration coefficient becomes the VP (permeability of capillary walls) multiplied by the surface area to volume ratio.

At a specific time point $t = t_0$, to analyze only the spatial characteristics of the fluid pressure, we can write

$$\frac{d^2p}{dR^2} + \frac{2}{R}\frac{dp}{dR} = \frac{\chi}{k}p - W, \qquad [11]$$

where $W = \frac{1}{H_A}\frac{dp}{dt} + \frac{1}{H_A}\frac{dQ}{dt}$ at $t_0$. Using the formula for the Laplace operator in spherical coordinates $\nabla^2 p = \frac{1}{R^2}\frac{d}{dR}[R^2\frac{dp}{dR}] = \frac{d^2p}{dR^2} + \frac{2}{R}\frac{dp}{dR}$, eq. (11) can be written as

$$\nabla^2 p = \frac{\chi}{k}p - W. \qquad [12]$$

Solving eq. (12) with boundary condition of zero fluid pressure at the periphery of the tumor, the equation for the fluid pressure can be written as (6, 7)

$$p(R) = \Psi\left(1 - \frac{\sinh(\alpha \frac{R}{a})}{\frac{R}{a}\sinh\alpha}\right), \quad \text{where} \quad \alpha = a\sqrt{\frac{L_p}{k}\frac{S}{V}}. \qquad [13]$$

Here $\Psi = \frac{W}{\alpha^2}$ is a constant and related to the peak fluid pressure $P_0$ as $P_0 = \Psi(1 - \alpha\,\text{cosech}(\alpha))$ and $a$ is the radius of the tumor.

The parameter $\alpha$ can be estimated by fitting the fluid pressure data (estimated using eq. (9)) with the theoretical equation of fluid pressure (eq. (13)). Using the value of the radius of the tumor and the surface area to volume ratio of the capillary walls, it is possible to determine the ratio between the VP and IP by knowledge of $\alpha$.

**C. Ratio between peak IFP and effective vascular pressure.** From eq. (1) of the manuscript, the peak IFP at the center of the tumor can be written as

$$P_i(0) = P_{i0} = \lim_{R \to 0} P_e\left(1 - \frac{\sinh(\alpha R)}{R\sinh\alpha}\right)$$
$$= P_e(1 - \alpha\,\text{cosech}(\alpha)). \qquad [14]$$

Therefore, the ratio between the peak IFP and effective vascular pressure can be expressed as

$$\frac{P_{i0}}{P_e} = 1 - \alpha\,\text{cosech}(\alpha). \qquad [15]$$

**D. Fluid velocity.** The interstitial fluid stress can be written in terms of the fluid pressure as (8)

$$\sigma_f(R,t) = -p(R,t). \qquad [16]$$

The equation of the equilibrium for the solid-fluid mixture can be written as

$$\nabla \sigma_f(R,t) + \frac{1}{k}(v_s(R,t) - v_f(R,t)) = 0, \qquad [17]$$

where $v_s$ is solid velocity and $v_f$ is the fluid velocity. Using eq. (16), eq. (17) can be written as

$$-\nabla p(R,t) - \frac{1}{k} v_s^f(R,t) = 0. \qquad [18]$$

The parameter $v_s^f$ is the fluid velocity with respect to the solid inside the tumor and can be expressed as

$$v_s^f(R,t) = -k\nabla p(R,t). \qquad [19]$$

Based on eq. (19), the permeability-normalized fluid velocity with respect to the solid along the radial direction inside a tumor can be written as

$$v_R(R,t) = -\frac{dp(R,t)}{dR}. \qquad [20]$$

As the fluid pressure $p$ is the weighted version of IFP (eq. (13)). the fluid velocity $v_R$ is also the weighted version of the IFV.

**E. Fluid flow.** The equation, which relates the fluid velocity and the time derivative of the volumetric strain, can be written as (9, p 6)

$$\frac{\delta \epsilon}{\delta t} = -\nabla q(R,t), \qquad [21]$$

where

$$q(R,t) = \psi(v_s(R,t) - v_f(R,t)). \qquad [22]$$

Here $\psi$ is the porosity.

In the original work of Biot (10) and in a subsequent work (11), the above equation is written as

$$\frac{\delta \xi}{\delta t} = -\nabla q(R,t), \qquad [23]$$

where $\frac{\delta \xi}{\delta t}$ is the change in the fluid content, which is created by the net outflow of the fluid presented by the quantity $\nabla q(R,t)$. Therefore, we can write the equation for the fluid flow as

$$w = \frac{\delta \xi}{\delta t} = \frac{\delta \epsilon}{\delta t}. \qquad [24]$$

From eqs. (23) and (24), it is clear that the net fluid flow is the gradient of the fluid velocity in a poroelastography experiment multiplied by the porosity of the tumor. Similarly, the fluid flow inside the tumor can be computed as the product of the gradient of the IFV and the porosity of the tumor. Therefore, the fluid flow in a poroelastography experiment would be the fluid flow inside the tumor multiplied by a constant factor.

**F. Axial strain time constant .** Based on a single exponential approximation, the axial strain inside the tumor can be written as (12)

$$\epsilon_{zz}(R,\theta,t) = f(R,\theta)\exp\left(-\frac{t}{\tau}\right), \qquad [25]$$

where $f(R,\theta)$ denotes the spatial variation of the axial strain and $\tau$ is the TC of the axial strain temporal curve. In eq. (25), the symmetry of the axial strain with respect to the third coordinate $\phi$ is assumed. However, axial strain is not symmetric with respect to the coordinate $\theta$, as indicated by eq. (25). In eq. (25), $\tau$ can be defined as (4)

$$\tau = \frac{\Omega}{H_A k} + \frac{1}{H_A \chi}. \qquad [26]$$

Here, $\Omega$ is a constant. which depends on the volumetric weight of the pore fluid. Poisson's ratio and radius of the tumor. We see from eq. (26) that the value of the axial strain TC $\tau$ is inversely proportional to the values of the IP and VP, i.e., if IP and/or VP increase inside a tumor, $\tau$ decreases and vice versa.

## 2. Validation of the proposed technique

### A. Simulations.

***A.1. Finite element simulations.*** A schematic of the poroelastic sample containing a spherical inclusion used in the validation of the proposed theory in this paper is shown in Fig. S1 (A). The sample is assumed to be of cylindrical shape, whereas the inclusion inside the sample is of spherical shape. Because of the cylindrical and spherical symmetry of the sample and the inclusion. a 2D solution plane for this problem can be assumed as shown in Fig. S1 (B). The sample is compressed from the top, and the bottom side is fixed. Two frictionless compressor plates are used for holding up the sample and exert compression upon it.

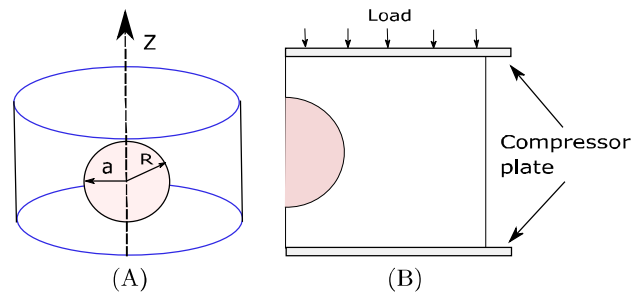

**Fig. S1.** (A) A schematic of a cylindrical sample of a poroelastic material with a spherical inclusion of radius $a$. The axial direction is along the $z$-axis. Inside the inclusion, $R$ indicates the radial direction. (B) The 2D solution plane for the three dimensional sample. The sample is compressed between two compressor plates. The compression is applied along the $z$ direction.

The commercial finite element simulation software ABAQUS, Abaqus Inc, Providence, RI, USA was used to validate the proposed technique developed in this paper. An 'effective stress' principle is used in ABAQUS (13), whereby the total stress acting at a point is assumed to be made up of the average pressure in the pore fluid and an 'effective/elastic stress' on the solid matrix. Both the inclusion and background of the samples used in the simulations were modeled as a linearly elastic, isotropic, permeable solid phase fully saturated with fluid.

Two samples with different mechanical properties were simulated in our study. An instantaneous load of 3000 Pa was applied to each sample and then kept constant while the sample was under compression. The interstitial permeability of the sample was assumed to be independent of the strain and void ratio. The mesh used to model the sample was CAX4RP with 63,801 elements in the solution plane. The dimension of the solution plane of the sample was 2 cm in radius and 4 cm in height. The radius of the inclusion was 0.3 cm. A zero fluid pressure boundary condition on the right hand side of the sample was imposed. The specific weight of the fluid was assumed to be 1 Nm$^{-3}$ to match the definitions of interstitial permeability in ABAQUS and in the developed model. Under the assumption of unit specific weight of the pore fluid, the hydraulic conductivity and permeability become equal (14). In ABAQUS, the microfiltration coefficient (vascular hydraulic conductivity or permeability multiplied by the surface area to volume ratio of the vessels) is modeled with the seepage coefficient. The void ratio used in all samples was 0.4. The time response of each sample was recorded for 60 second with a 0.3 sampling interval. The instantaneous load of 3000 Pa was applied in the first 0.01 second. This load was then kept constant for 60.01 s. Further details of the poroelastic simulation can be found in (15, 16).

The mechanical properties of the samples used in our simulations were chosen following (4, 7). In all cases, the Poisson's ratio was assumed to be 0.49 in the background (normal) tissue and 0.47 in the inclusion (tumor) (4). The Young's moduli of the tumor and normal tissue were assumed to be 32.78 and 97.02 kPa (4). The interstitial permeability of the normal tissue was always assumed to be 1000 times higher than the interstitial permeability of the tumor. Similar values of interstitial permeability contrast between tumor and surrounding tissue have been previously considered in the literature (17) (supplementary p. 17). In the first sample, the interstitial and vascular permeability (hydraulic conductivity) has comparable effect, while in the other sample the vascular permeability has much higher effect than the interstitial permeability. The effect of vascular permeability can be measured by value of $\chi$ and of interstitial permeability can be measured by $\frac{k}{a^2}$. Table S1 provides a detailed description of the samples used for the simulation study. In this table, the parameters without subscript are pertinent to the inclusion (tumor) and with subscript b are pertinent to the background (normal tissue).

**Table S1. Mechanical parameters of samples A and B**

| Sample name | $E_b$ (kPa) | $E$ (kPa) | $\nu_b$ | $\nu$ | $k_b$ (m$^4$N$^{-1}$s$^{-1}$) | $k$ (m$^4$N$^{-1}$s$^{-1}$) | $\chi_b$ (Pa s)$^{-1}$ | $\chi$ (Pa s)$^{-1}$ |
|---|---|---|---|---|---|---|---|---|
| A | 32.78 | 97.02 | 0.49 | 0.47 | $3.19 \times 10^{-9}$ | $3.19 \times 10^{-12}$ | $1.89 \times 10^{-9}$ | $5.67 \times 10^{-8}$ |
| B | 32.78 | 97.02 | 0.49 | 0.47 | $3.10 \times 10^{-11}$ | $3.10 \times 10^{-14}$ | $1.89 \times 10^{-7}$ | $5.67 \times 10^{-7}$ |

To estimate the fluid pressure inside the tumor from finite element method (FEM) as well as ultrasound simulation data, we require to estimate the Young's modulus and Poisson's ratio of the tumor, which were estimated using method described in Ref. (18).

To determine the value of $\alpha$ in samples A and B, we fit the equation of the fluid pressure (eq. (13)) on the fluid pressure estimated from FEM strain data (from the center to a radial direction) of samples A and B at time point of 3 s and 10 s, respectively. We used 'Levenberg Marquardt' algorithm in Matlab (Matlab Inc, Natick, MA) for the curve fitting purpose. We determined $\alpha$ for all samples in ultrasound simulations in the same manner by fitting the equation of the fluid pressure (eq. (13)) on the fluid pressure data estimated from simulation RF data. A smoothing filter is of length 5 is applied on the fluid pressure data before estimating the $\alpha$.

**A.2. Ultrasound simulations.** The simulated pre- and post-compression temporal ultrasound radio frequency (RF) data were generated from the mechanical displacements from FEM using a convolution model (19). Bilinear interpolation was performed on the input mechanical displacement data prior to the computation of the simulated RF frames (20). The simulated ultrasound transducer had 128 elements, frequency bandwidth between $5 - 14$ MHz, a 6.6 MHz center frequency, and 50% fractional bandwidth at $-6$ dB. The transducer's beamwidth was assumed to be dependent on the wavelength and to be approximately 1 mm at 6.6 MHz (21). The sampling frequency was set at 40 MHz and Gaussian noise was added to set the SNR at 40 dB. To compute the axial and lateral strain elastograms from ultrasound pre- and post-compressed RF data, the method proposed in Ref. (22) has been used.

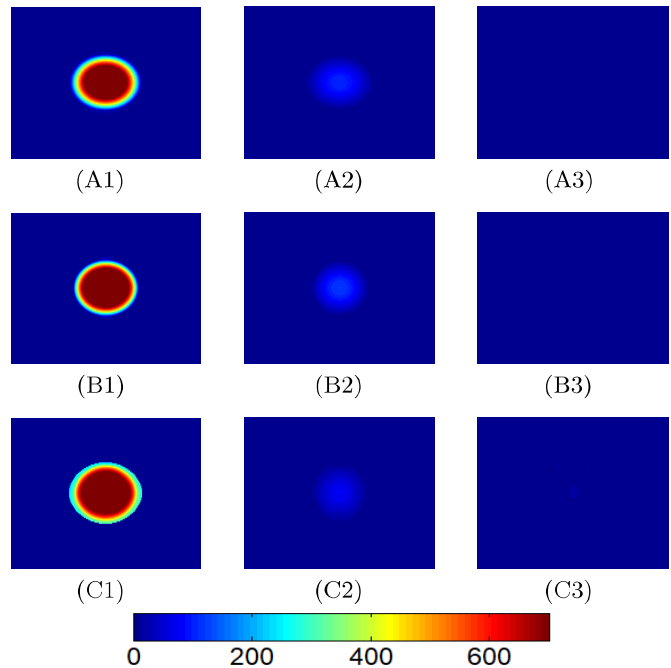

**Fig. S2.** Fluid pressure from FEM and fluid pressure estimated by the proposed method using the FEM axial and lateral strains at time points of 4 s, 60 s and 120 s for sample A are shown in (A1-A3) and in (B1-B3) respectively. Fluid pressure estimated by the proposed method using ultrasound simulated axial and lateral strains at time points of 4 s, 60 s and 120 s for sample A are shown in (C1-C3). The unit of fluid pressure is Pa.

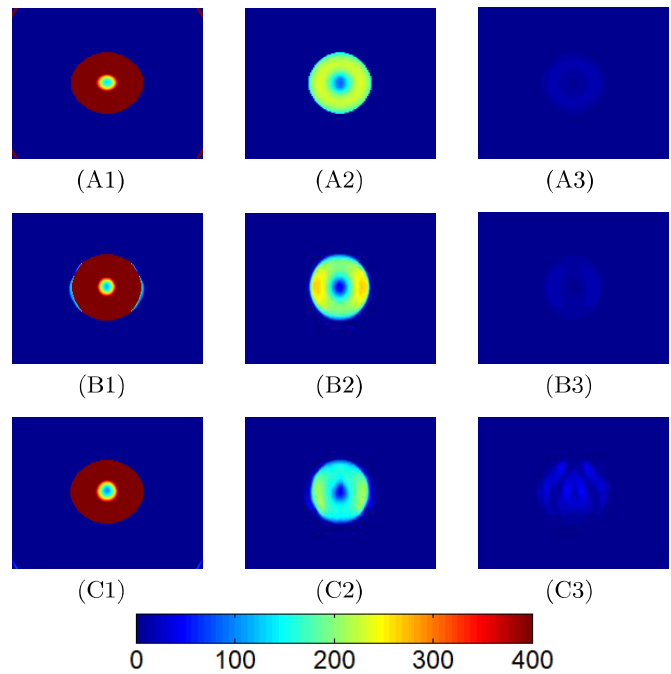

**Fig. S3.** Fluid velocity from FEM and fluid velocity estimated by the proposed method using the FEM axial and lateral strains at time points of $4$ s, $60$ s and $120$ s for sample A are shown in (A1-A3) and in (B1-B3) respectively. Fluid velocity estimated by the proposed method using ultrasound simulated axial and lateral strains at time points of $4$ s, $60$ s and $120$ s for sample A are shown in (C1-C3). The unit of fluid velocity is Pa cm$^{-1}$.

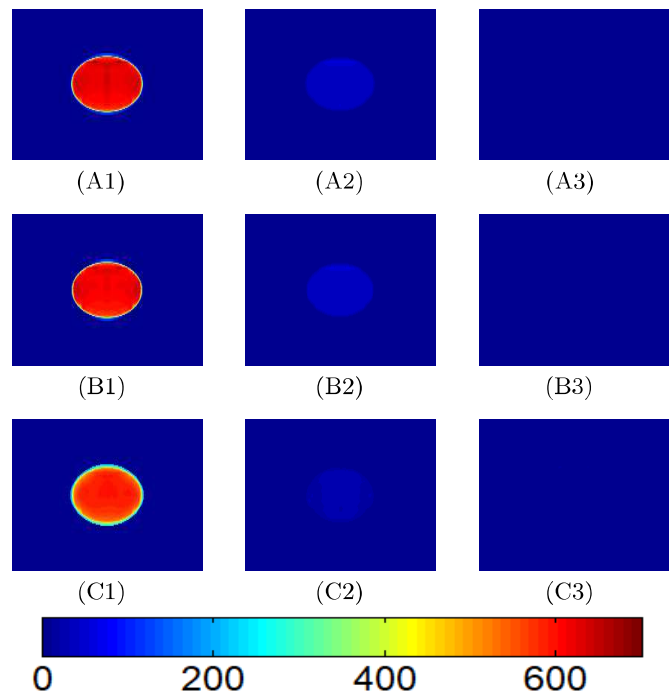

**Fig. S4.** Fluid pressure from FEM and fluid pressure estimated by the proposed method using the FEM axial and lateral strains at time points of $4$ s, $60$ s and $120$ s for sample B are shown in (A1-A3) and in (B1-B3) respectively. Fluid pressure estimated by the proposed method using ultrasound simulated axial and lateral strains at time points of $4$ s, $60$ s and $120$ s for sample B are shown in (C1-C3). The unit of fluid pressure is Pa.

**B. Results.** To validate the proposed method, we show the fluid pressure and fluid velocity from FEM along with these field parameters estimated using the proposed method from both FEM and ultrasound axial and lateral strains in Figs. S2 and S3 for sample A. For sample B, the same parameters from FEM, estimated by the proposed method from FEM and ultrasound simulated strain data have been shown in Figs. S4 and S5. From Figs. S2, S3, S4 and S5, we see that the fluid pressure and fluid velocity from the FEM and estimated by the proposed method match well.

The fitted curve with equation of $\alpha$ (eq. (13)) along the radial profile of fluid pressure estimated from FEM strain data of samples A and B using the proposed technique is shown in Fig. S6 (A) and (B). The true $\alpha$ computed from the mechanical parameters and the radius of the tumor and the estimated $\alpha$ match with less than 10% error. The ratio of the microfiltration coefficient and interstitial permeability ($\frac{L_p S}{kV}$) can be estimated from values of $\alpha$, which are 19228.4 and 16181847.1 m in samples A and B, which match well with the mechanical parameters (see Table S1) considered in the simulation. Similar values of the $\alpha$ and the ratio of microfiltration coefficient and interstitial permeability are obtained for the ultrasound simulation data and the error in estimation of these parameters remain below 20% in both of the samples, which are shown in Fig. S6 (C) and (D).

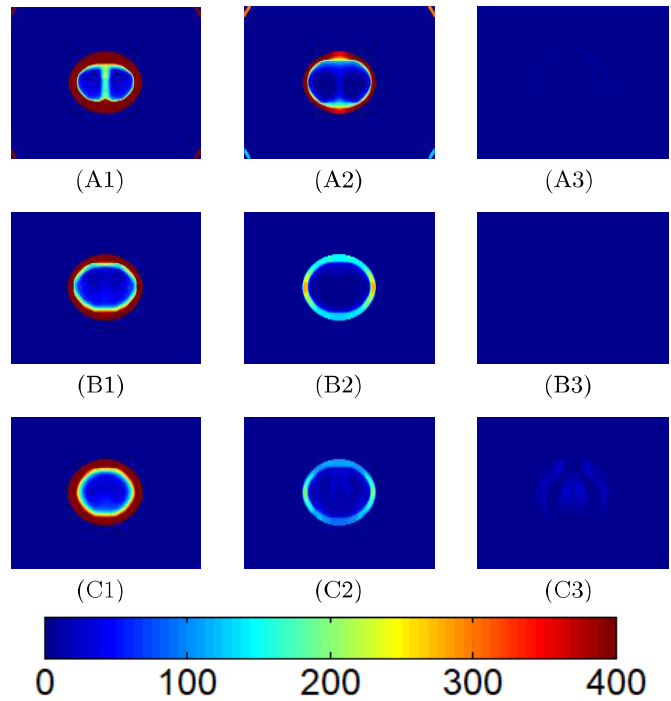

**Fig. S5.** Fluid velocity from FEM and fluid velocity estimated by the proposed method using the FEM axial and lateral strains at time points of $4$ s, $60$ s and $120$ s for sample A are shown in (A1-A3) and in (B1-B3) respectively. Fluid velocity estimated by the proposed method using ultrasound simulated axial and lateral strains at time points of $4$ s, $60$ s and $120$ s for sample B are shown in (C1-C3). The unit of fluid velocity is Pa cm$^{-1}$.

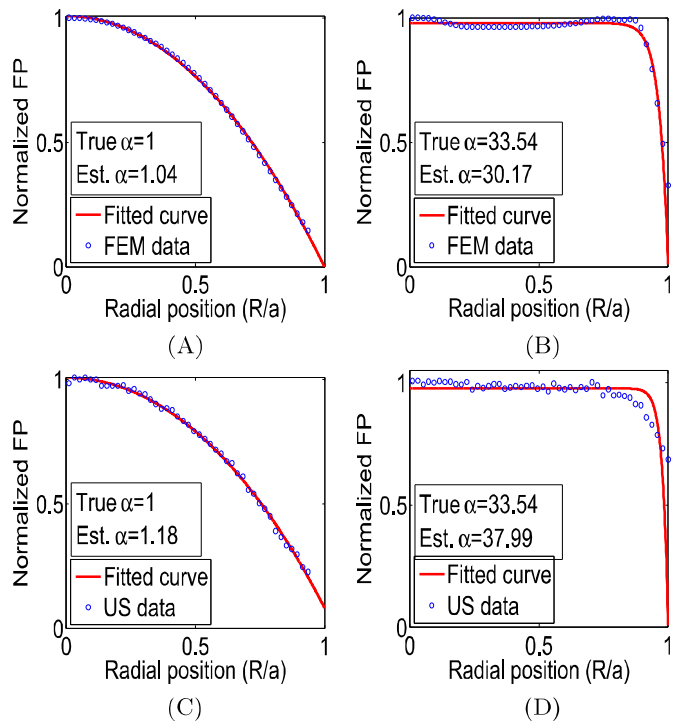

**Fig. S6.** Curve fitting for computing α from radial profile of fluid pressure (FP) estimated from FEM strain data of sample A (A) and sample B (B). Curve fitting for computing α from radial profile of fluid pressure estimated from ultrasound simulated (US) strain data of sample A (C) and sample B (D).